\documentclass[sigconf, balance=false, article]{acmart}
\usepackage[T1]{fontenc}
\usepackage[edges]{forest}
\usepackage{amsthm}
\usepackage{bbding}
\usepackage{listings}
\usepackage{multicol}
\usepackage[htt]{hyphenat}
\usepackage{subcaption}
\usepackage{xspace}
\usepackage{balance}
\usepackage{fancybox} %
\usepackage{xcolor}    %
\usepackage{url}
\usepackage{hyperref}
\usepackage[hyphenbreaks]{breakurl}

\lstdefinestyle{customstyle}{
  backgroundcolor=\color{gray!20}, 
  basicstyle=\ttfamily, 
  frame=single, 
  captionpos=b, 
  breaklines=true, 
  numbers=left, 
  numberstyle=\tiny\color{gray}, 
  commentstyle=\color{black!40}, 
  morecomment=[l]{//}, 
  escapechar=\%,
  xleftmargin=3.4pt,
  xrightmargin=3.4pt,
}

\newcommand{\GHilight}{\makebox[0pt][l]{\color{green}\rule[-4pt]{\linewidth}{10pt}}}

\AtBeginDocument{%
  \providecommand\BibTeX{{%
    \normalfont B\kern-0.5em{\scshape i\kern-0.25em b}\kern-0.8em\TeX}}}

\copyrightyear{2025}
\acmYear{2025}
\setcopyright{cc}
\setcctype{by-nc-sa}
\acmConference[SCORED '25]{Proceedings of the 2025 Workshop on Software Supply Chain Offensive Research and Ecosystem Defenses}{October 13--17, 2025}{Taipei, Taiwan}
\acmBooktitle{Proceedings of the 2025 Workshop on Software Supply Chain Offensive Research and Ecosystem Defenses (SCORED '25), October 13--17, 2025, Taipei, Taiwan}\acmDOI{10.1145/3733827.3765523}
\acmISBN{979-8-4007-1915-8/2025/10}

\newcommand{\TODO}[1]{\textcolor{red}{#1}\GenericWarning{}{LaTeX Warning: TODO: #1}}\newcommand\todo\TODO
\newcommand{\attackname}{Maven-Hijack\xspace}

\begin{document}

\title{\attackname: Software Supply Chain Attack \\
Exploiting Packaging Order}


\setcopyright{none}

\author{Frank Reyes}
\affiliation{
  \institution{KTH Royal Institute of Technology}
  \city{Stockholm}
  \country{Sweden}
}
\email{frankrg@kth.se}

\author{Federico Bono}
\affiliation{
  \institution{KTH Royal Institute of Technology}
  \city{Stockholm}
  \country{Sweden}
}
\email{fbono@kth.se}

\author{Aman Sharma}
\affiliation{
  \institution{KTH Royal Institute of Technology}
  \city{Stockholm}
  \country{Sweden}
}
\email{amansha@kth.se}

\author{Benoit Baudry}
\affiliation{
  \institution{Universtité de Montr\'eal}
  \city{Montr\'eal}
  \country{Canada}
}
\email{benoit.baudry@umontreal.ca}

\author{Martin Monperrus}
\affiliation{
  \institution{KTH Royal Institute of Technology}
  \city{Stockholm}
  \country{Sweden}
}
\email{monperrus@kth.se}

\begin{abstract}


Java projects frequently rely on package managers such as Maven to manage complex webs of external dependencies.
While these tools streamline development, they also introduce subtle risks to the software supply chain.
In this paper, we present \attackname, a novel attack that exploits the order in which Maven packages dependencies and the way the Java Virtual Machine resolves classes at runtime.
By injecting a malicious class with the same fully qualified name as a legitimate one into a dependency that is packaged earlier, an attacker can silently override core application behavior without modifying the main codebase or library names.
We demonstrate the real-world feasibility of this attack by compromising the Corona-Warn-App, a widely used open-source COVID-19 contact tracing system, and gaining control over its database connection logic.
We evaluate three mitigation strategies, such as sealed JARs, Java Modules, and the Maven Enforcer plugin.
Our results show that, while Java Modules offer strong protection, the Maven Enforcer plugin with duplicate class detection provides the most practical and effective defense for current Java projects.
These findings highlight the urgent need for improved safeguards in Java's build and dependency management processes to prevent stealthy supply chain attacks.

\end{abstract}


\keywords{Software Supply Chain Attack, Java, Maven, Gradle, Namespace, Class Hijacking}

\maketitle

\begin{figure*}[!ht]
  \centering
  \includegraphics[width=0.95\linewidth]{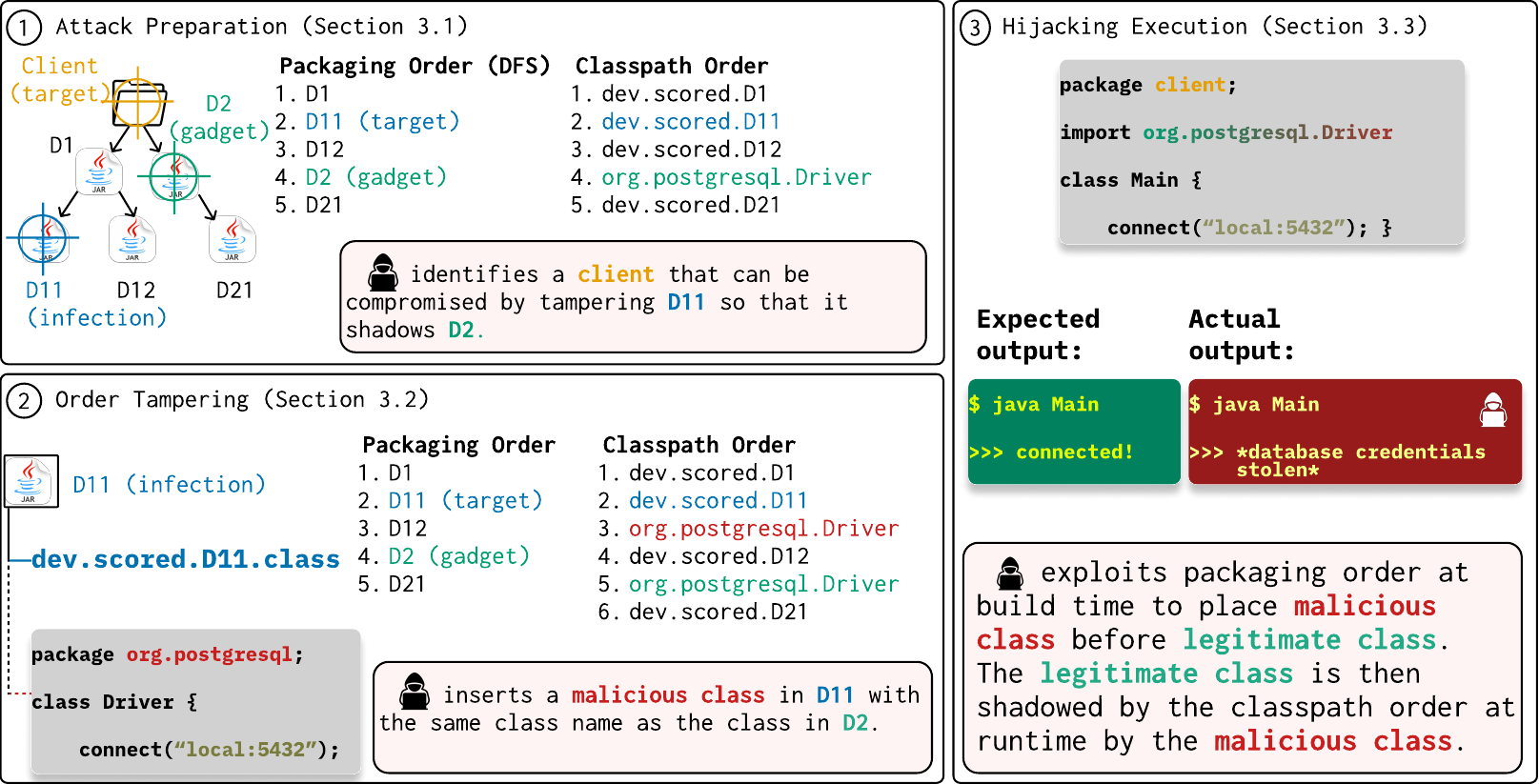}
  \caption{Overview of the \attackname attack.}
  \label{fig:attack}
\end{figure*}

\section{Introduction}

Modern software development relies heavily on automated build tools and extensive dependency ecosystems.
Package managers such as Maven and Gradle simplify the integration of third-party libraries, but they also introduce new risks to the software supply chain.
As dependency trees grow deeper and more complex, attackers are increasingly able to exploit blind spots in the build process to introduce malicious artifacts.

Recent attacks, such as dependency confusion~\citep{ladisa_taxonomy_2022} and typosquatting~\citep{liu_exploring_2022}, illustrate how ambiguous naming or domain hijacking allow adversaries to inject untrusted code into otherwise secure environments.
For example, the MavenGate attack exploited ownership assumptions in Maven's groupId namespace~\citep{lakshmanan_mavengate_2024}, enabling attackers to publish malicious versions of previously trusted libraries by re-registering expired domains.
These incidents highlight a critical weakness: build systems often lack sufficient checks to verify the integrity and origin of the code they incorporate.

In this paper, we introduce \attackname, a new class of software supply chain attack that exploits the order in which dependencies are packaged and the way the Java Virtual Machine resolves classes at runtime.
By injecting a class with the same fully qualified name as one in a direct dependency (the gadget dependency) into another dependency (the infection dependency), an attacker can silently override core application behavior.
Unlike prior attacks, \attackname does not require renaming artifacts or overriding versions; it relies solely on packaging order and classpath search behavior.

At build time, Maven packages project artifacts by traversing the dependency tree in depth-first order. Then, at runtime, the Java class loader performs a sequential scan of the classpath and loads the first match of a fully qualified class name.
As a result, classes from certain dependencies may take precedence over others in the final artifact.
A malicious class placed earlier in this sequence will overshadow any legitimate class with the same name that is packaged later~\citep{wang_will_2022}.

We demonstrate the feasibility of \attackname through a detailed proof of concept targeting the Corona-Warn-App, a widely used open-source COVID-19 contact tracing system.
We inject a malicious class that hijacks the database initialization process into a transitive dependency (infection) that appears early in the build order, and manage to gain full control over the database connection logic without altering any direct dependencies or the application source code.

To assess potential defenses, we evaluate several mitigation strategies.
Sealed JARs block conflicting classes at runtime but can be bypassed by copying entire packages.
Java Modules provide strong protection by detecting class duplication during compilation, though their adoption remains very low.
The Maven Enforcer plugin, when configured with banDuplicateClasses, stops the build if duplicate classes are found, offering both strong protection and practical applicability.
Among these, we find Maven Enforcer to be the most effective and actionable defense currently available.
We discuss these mitigations in detail in \autoref{sec:mitigations}.

Our contributions are as follows:
\begin{enumerate}
    \item We define \attackname, a novel class of software supply chain attack that leverages order manipulation to hijack runtime behavior. 
    \item A working proof of concept implemented on a real-world application, demonstrating the feasibility of \attackname.
    The proof of concept is publicly available at the repository \url{https://github.com/chains-project/maven-class-hijack-poc/}.
    \item An extension of the analysis to the Gradle ecosystem, analyzing the conditions under which similar class hijacking attacks are possible.
    \item An evaluation of existing mitigation strategies, including stricter packaging policies, sealed jars, and enforced modularization, assessing their effectiveness against \attackname.
\end{enumerate}

\section{Background}

We introduce two key steps that occur when developers build a Java project with Maven: Maven packages an artifact for the project, and the Java classloader loads classes from that artifact.
The design decisions made in Maven and the JVM for these two steps lead to the feasibility of \attackname, a software supply chain attack on the built project.

\begin{figure}[h]
  \centering
  \begin{forest}
    for tree={grow=south, s sep=1em, shape=rectangle, rounded corners, 
              draw, align=center, top color=white, bottom color=blue!20}
    [Project
      [D1, name=D1, tikz={\node [draw,red,fit=(D1) (D2), label=left:Direct Dependencies] {};}
        [D11, name=D11, tikz={\node [draw,red,fit=(D11) (D111) (D112) (D21) (D22) (D211) (D221),  label=left:Transitive Dependencies] {};}
          [D111, name=D111]
          [D112, name=D112]
        ]
      ]
      [D2, name=D2
        [D21, name=D21
          [D211, name=D211]
        ]
        [D22, name=D22
          [D221, name=D221]
        ]
      ]
    ]
  \end{forest}
  \caption{Resolved dependency tree of a sample Maven project showing its direct and transitive dependencies.}
  \label{fig:dependency-tree}
\end{figure}

\subsection{Packaging in Maven}
\label{sec:artifact-packaging}

Packaging is a process at build time that bundles the compiled source code of the project and all the resolved dependencies into a single artifact.
In the context of Maven, this single file is a compressed archive and is called an uber jar.
This jar file consists of the classfiles of the project, together with classfiles or jars of the dependencies.
Depending upon the maven plugin used to package the project, the dependencies are either included as classfiles in the uber jar~\footnote{\url{https://maven.apache.org/plugins/maven-shade-plugin/}} or as nested jar files in the uber jar~\footnote{\url{https://docs.spring.io/spring-boot/specification/executablejar/nested-jars.html}}.
The order of files in the compressed archive is determined by iterating through the dependency tree of the project in Depth First Search (DFS) Order~\footnote{\url{https://github.com/apache/maven-resolver/blob/master/maven-resolver-util/src/main/java/org/eclipse/aether/util/graph/visitor/TreeDependencyVisitor.java}}.
The first node is the classes of the project itself, and the subsequent nodes are the classes of the dependencies.
For example, the DFS order of the dependency tree in \autoref{fig:dependency-tree} is \texttt{Project}, \texttt{D1}, \texttt{D11}, \texttt{D111}, \texttt{D112}, \texttt{D2}, \texttt{D21}, \texttt{D211}, \texttt{D22}, \texttt{D221}.

\subsection{Class loading in Java}

Class loading in Java is the process of locating
classfiles based on their fully qualified names. This process occurs at runtime.
Whenever a particular class is invoked during the execution of a Java application, the Java runtime environment locates the class in the classpath, which consists of paths to Jar files or to classfiles (on the file system or URLs).
To locate the class in the classpath, Java employs a linear search mechanism over the classpath.
This means that it checks each entry in the classpath, one by one until it finds the class whose fully qualified name is equal to the one that has been invoked~\cite{wang_will_2022}.
If the path is a classfile, it checks its name before loading it.
If the path is a jar file, it recursively extracts the classfiles from the jar file and checks their names.
This linear search mechanism is the key to the \attackname attack, as it allows an attacker to hijack a class by placing a malicious class with the same name as a legitimate class earlier in the classpath.
We illustrate this software supply chain attack in the next section.

\section{Attack Concept}
\label{sec:concept}
In this section, we present an overview of the \attackname attack as shown in \autoref{fig:attack}.
The attack is a three step process of preparation, order tampering, and finally, hijacking the legitimate class with a malicious Java class.
The success of the attack is the result of packaging process by Maven at build time and then the class loading mechanism of Java at runtime.

\subsection{Attack Preparation}

In the first step, the attacker identifies a target application to compromise.
The \textbf{target application} is a project that has two weaknesses in its dependency tree: a `gadget' dependency and an `infection' dependency, defined as follows. A code gadget usually refers to a snippet of code misused to exploit an application~\cite{sayar_-depth_2023}. Similarly, we introduce the concept of \textbf{gadget dependency} to refer to a dependency, which identifiers can be reused to compromise the classpath of an application. 
An \textbf{infection dependency} is a weak link in the dependency tree of the target application where the attacker can inject a malicious class.
It can be a dependency that is not maintained for a long time or has very few commits, yet many applications use it.
This kind of dependency can be compromised by the attacker using multiple social engineering techniques, such as account compromise or infiltrating the maintainer team ~\cite{siadati_devphish_2024}.

Both a gadget dependency and an infection dependency are necessary to compromise the target application.
The gadget dependency contains the class that the attacker will shadow with a malicious one.
The attacker will inject the malicious class in the infection dependency.
Eventually, the attacker will be able to ensure that the malicious class is executed instead of the legitimate one.
In our threat model, the attacker must be able to modify the Pom file of a project.
There are different possibilities for that:
gain publishing rights over the infection dependency, compromise publishing credentials, or reclaim abandoned projects.
At the end, all those options yield that same result: a malicious is centrally pushed and accepted as legitimate by Maven’s infrastructure.

\subsection{Adding a Malicious Class in the Classpath}

After taking control of the infection dependency via social engineering, the attacker injects malicious code in the infection dependency.
The malicious code is a class that has the same fully qualified name as the legitimate class in the gadget dependency.
This malicious version is crafted to perform malicious actions, such as stealing sensitive data or executing arbitrary code.
The malicious class is then bundled together with the infection dependency and uploaded to a public repository such as Maven Central.
Next time the target application is built, the malicious class is included in the artifact after packaging, per the order of the packaging algorithm, see \autoref{sec:artifact-packaging}.
At this point, this artifact to be executed in production is compromised.

\subsection{Hijacking Execution at Runtime}

In the final step, the attacker hijacks the class loading mechanism of Java.
When the victim's application is executed, the Java classloader searches for the legitimate class, from gadget dependency, in the classpath per the order of elements in the classpath and Jar files.
Since the infection dependency is included before the gadget dependency, the classloader finds the malicious class first, loads it instead of the legitimate one, and subsequent execution of the compromised class is fully controlled by the attacker.

Overall, the victim's application is compromised via its dependencies, exploiting subtleties of packaging and classloading of the platform under study, Java, making the attack a sophisticated software supply chain attack.

\section{Proof of Concept}
\label{sec:proof}

We have implemented the attack in a proof of concept to demonstrate the practical feasibility of \attackname.
It is based on a real-world open-source project, the Corona-Warn-App\footnote{\url{https://github.com/corona-warn-app/cwa-server}}. 

Corona-Warn-App is Germany's coronavirus tracking application, used during the 2020 pandemic to perform contact tracing.
The backend services are written using the popular Java framework, Spring Boot.

\begin{figure}[h]
  \centering
  \begin{forest}
    for tree={grow=south, s sep=1em, shape=rectangle, rounded corners, 
              draw, align=center, top color=white, bottom color=blue!20}
    [cwa-server, name=cwa-server, tikz={\node[draw, yellow, fit=(cwa-server), label=above:\shortstack{target application}] {};}
      [...]
      [everit-json-schema, name=D11, tikz={\node [draw,blue,fit=(D11), label=below:\shortstack{infection dependency\\is defined before the gadget}] {};}
        ]
      [...]
      [postgresql, name=database, tikz={\node [draw,green,fit=(database), label=below:\shortstack{gadget}] {};}]
    ]
  \end{forest}
  \caption{Truncated dependency tree for Corona-Warn-App backend service, from the parent pom.xml file}
  \label{fig:cwa-tree}
\end{figure}

Corona-Warn-App is composed of multiple microservices, each with its own dependencies but sharing a common parent \texttt{pom.xml} file.
Being able to compromise the parent dependency tree means that any service will also be vulnerable to the attack.

In our proof-of-concept, we identify \texttt{everit-json-schema} as the infection dependency, which appears in the dependency tree before the gadget dependency, the PostgreSQL JDBC driver identified by the artifact \texttt{org.postgresql\allowbreak{}:postgresql}.
This is a powerful gadget because database connections are initialized at startup, providing the attackers with guarantees that the payload will be executed. 

\autoref{fig:cwa-tree} shows an overview of the target application's dependency tree, illustrating how the infection dependency precedes the gadget dependency. 
This ordering is critical to ensure that the malicious class is encountered first during classpath resolution.
Consequently, an attacker who can include a malicious class in  \texttt{everit-json-schema}, or in any of its transitive dependencies, can control the implementation for any class from the \texttt{postgresql} SDK, effectively overtaking the database connection layer.

We simulate a library takeover scenario by publishing a compromised version of \texttt{everit-json-schema}. The compromised version does not alter any existing class in the original library.
Instead, it adds a single malicious dependency, which contains a malicious implementation of the class \texttt{\path{org.postgresql.Driver}} the same fully qualified name as the legitimate one found in the PostgreSQL SDK.

When the backend service starts, Spring Boot attempts to establish a database connection using the driver class.
The malicious class is crafted to silently exfiltrate sensitive information from the client during runtime.
The classloader loads the malicious \texttt{org.postgresql.Driver} from the compromised \texttt{everit-json-schema} dependency instead of the legitimate one, because of the logic of the attack.
As a result, the attacker's custom logic is executed transparently, enabling operations such as reading credentials, intercepting queries, or leaking confidential data from the application or the machine it is running on.

The complete proof-of-concept demonstrates a successful realization of the attack stages described in \autoref{sec:concept}.
The prototype and scripts are available on our GitHub repository \href{https://github.com/chains-project/maven-class-hijack-poc/}{\path{https://github.com/chains-project/maven-class-hijack-poc/}}.

\section{Mitigations}
\label{sec:mitigations}

We now present three different strategies to mitigate \attackname.

\subsection{Sealed JARs}
A \href{https://docs.oracle.com/javase%2Ftutorial%2F/deployment/jar/sealman.html}{sealed JAR}~\footnote{\url{https://docs.oracle.com/javase\%2Ftutorial\%2F/deployment\\/jar/sealman.html}} enforces that all classes belonging to a specific Java package must be loaded from the same archive.
This provides a defense against \attackname by preventing the JVM from mixing classes of the same package from different JARs.
In \autoref{fig:attack}, the legitimate \texttt{\path{org.postgresql.Driver}} class is defined in dependency \texttt{D2} (gadget).
If \texttt{D2} is published as a sealed JAR, then the malicious version of \texttt{\path{org.postgresql.Driver}} in \texttt{D11}(infection) will violate this restriction.

\autoref{lst:sealed} shows the modified \texttt{MANIFEST.MF} for \texttt{D2} to seal all packages in the jar including the \texttt{org.postgresql} package.
When the JVM attempts to load the malicious class \texttt{\path{org.postgresql.Driver}} from \texttt{D11}, a SecurityException is thrown at runtime as JVM only allows loading of classes in \texttt{org.postgresql} package from the gadget dependency \texttt{D2} and not from the infection dependency \texttt{D11}.
This prevents loading and execution of the malicious class.

However, this defense can be bypassed if the attacker includes a fully self-contained copy of the original package in \texttt{D11}, including all the classes that would be expected by the victim.
This is because the JVM will load all classes from the infection dependency \texttt{D11} and classpath search will not look inside the gadget dependency \texttt{D2}.
Although this replication is non-trivial, it is achievable by a powerful attacker.
In summary, sealed JARs strengthen integrity checks for individual packages but can be bypassed.

\begin{lstlisting}[backgroundcolor=\color{gray!10}, caption={Modified MANIFEST.MF for D2 to seal the all packages}, label=lst:sealed, style=customstyle, numbers=none]
  // postgresql-42.7.7.jar/MANIFEST.MF
  Bundle-Description: Java JDBC driver for PostgreSQL database
  ...
%\GHilight%+ Sealed: true
\end{lstlisting}

\subsection{Java Modules}

With Java 9, \href{https://openjdk.org/projects/jigsaw/}{\texttt{Project Jigsaw}}~\footnote{https://openjdk.org/projects/jigsaw/} was introduced into the JDK, adding the concept of modularity to the Java ecosystem.
Java Modules facilitate the maintenance of large projects and libraries while enhancing the security of the Java SE Platform.
A modularized application is immune to \texttt{\attackname} because a package collision results in an automatic compilation failure.
There are two ways to mitigate the attack leveraging Java Modules.
1) The target application can be defined as a module and all direct and transitive dependencies can be imported as modules.
2) The target application along with all direct and transitive dependencies can be defined as a module.

\begin{lstlisting}[backgroundcolor=\color{gray!10}, caption={Example of module-info.java to mitigate the attack}, label=lst:module,style=customstyle, numbers=none]
  module victim {
    // direct dependencies
    requires D1;
    requires org.postgresql.jdbc;
    // transitive dependencies
    requires D11; 
    requires D12;
    // Java platform modules
    requires java.sql;
  }
\end{lstlisting}

The first approach requires creating \texttt{module-info.java} file for the target application and then importing all direct and transitive dependencies as modules as shown in \autoref{lst:module}.
\texttt{D11} is the dependency that contains the malicious class.
If this is not imported as a module, the Java module system will not detect the conflict at build time.
This makes the approach not effective as the developer of the target application needs to be aware of all the transitive dependencies, even though they may not be directly used in the application.

\begin{lstlisting}[backgroundcolor=\color{gray!10}, caption={Example of module-info.java to mitigate the attack}, label=lst:module2,style=customstyle, numbers=none, float]
  module victim {
    // direct dependencies
    requires D1;
    requires org.postgresql.jdbc;
  }
  module D1 {
    // direct dependencies
    requires D11;
    requires D12;
  }
  // omitted Java platform modules for brevity
\end{lstlisting}

The second approach requires declaration of modules for each dependency and the target application as shown in \autoref{lst:module2}.
This approach is more effective as it allows the developer to only import the dependencies that are directly used in the application.
However, Java modules are not widely used in practice.
Bot et al.~\cite{bot_uncovering_2023} analyzed over 473,000 artifacts in Maven Central and found that only 1.69\% included a module-info.java file, confirming the low adoption of the module system and its limited mitigation coverage.

\begin{lstlisting}[backgroundcolor=\color{gray!10}, caption={Compilation error triggered by conflicting package in Java Modules}, label=lst:module-log,style=customstyle, numbers=none]
  [ERROR] the victim module reads package org.postgresql from both org.postgresql.jdbc and D1
\end{lstlisting}

In both approaches, the Java module system will detect that package \texttt{org.postgresql} is defined in multiple dependencies - gadget and infection dependencies and fail the build process with a compilation error as shown in \autoref{lst:module-log}.

\subsection{Maven Enforcer Plugin}

In Java projects built with Maven, developers can use the \texttt{maven-enforcer-plugin}~\footnote{https://maven.apache.org/enforcer/maven-enforcer-plugin/}, which includes a \href{https://www.mojohaus.org/extra-enforcer-rules/banDuplicateClasses.html}{\texttt{banDuplicateClasses}}~\footnote{https://www.mojohaus.org/extra-enforcerrules/banDuplicateClasses.html} option. 
When the \texttt{banDuplicateClasses} option is enabled, Maven fails the build process if a class collision is detected. Based on \autoref{fig:attack}, the plugin detects the overlap between \texttt{D11} (infection) and \texttt{D2} (gadget) due to the same class name \texttt{org.postgresql.Driver}.
\autoref{lst:enforce} shows the overlapping dependencies and conflicting classes:

\begin{lstlisting}[backgroundcolor=\color{gray!10}, caption={Example of maven-enforce plugin output to mitigate the attack}, label=lst:enforce,style=customstyle, numbers=none]
[ERROR] Found in:
[ERROR]  dev.scored:D1:jar:1.0.0:compile
[ERROR]  org.postgresql:postgresql:jar:42.6.0:compile
[ERROR] Duplicate classes:
[ERROR]  org/postgresql/Driver.class
\end{lstlisting}
    
While effective at detecting class shadowing, this rule may also flag legitimate duplicate classes for example, when libraries split packages across multiple artifacts.
Developers should assess whether triggering a compilation failure is acceptable in such cases, carefully weighing potential false positives against the intended security improvements.


\begin{center}
\colorbox{gray!10}{
\parbox{0.95\columnwidth}{%
\normalsize
    We have identified 3 mitigation strategies: 1 resulting in a runtime error and 2 resulting with a build error.    
    Among the mitigation strategies evaluated, \textbf{Maven Enforcer Plugin} provides the most principled robust protection against the \attackname attack.
    Unlike Sealed Jars, the attack is mitigated by the plugin at build time, much earlier than at runtime.
    Unlike Java Modules, the plugin is independent of how the dependencies are packaged.
    Maven Enforcer Plugin only requires the developer to enable the plugin.
}%
}
\end{center}
\section{Discussion}

The success of \attackname depends on how build systems resolve dependencies and package artifacts. We first examine various Maven packaging plugins, which differ in handling class conflicts and dependency order, affecting \attackname attack feasibility.
We then assess how Gradle's distinct design choices impact the practicality of launching the same attack.

\subsection{Impact of Packaging Plugins}

\begin{table}[h]
  \centering
  \begin{tabular}{|l|c|}
  \hline
  \textbf{Packaging Plugin} & \textbf{Attack Successful?} \\
  \hline
  \texttt{\href{https://maven.apache.org/plugins/maven-jar-plugin/}{maven-jar-plugin}} & yes \\
  \texttt{\href{https://docs.spring.io/spring-boot/maven-plugin/index.html}{spring-boot-maven-plugin}} & yes \\
  \texttt{\href{https://maven.apache.org/plugins/maven-shade-plugin/}{maven-shade-plugin}} & yes* \\
  \texttt{\href{https://quarkus.io/guides/quarkus-maven-plugin}{quarkus-maven-plugin}} & yes* \\
  \texttt{\href{https://felix.apache.org/documentation/subprojects/apache-felix-maven-bundle-plugin-bnd.html}{maven-bundle-plugin}} & yes\dag \\
  \texttt{\href{https://maven.apache.org/plugins/maven-assembly-plugin/}{maven-assembly-plugin}} & sometimes \\
  \hline
  \end{tabular}
  \caption{Attack feasibility across different Maven plugins that can be used for packaging.
  * indicates that the plugin emits warnings during packaging, but the attack still succeeds.
  \dag~indicates that the plugin requires the infection dependency to appear after the gadget dependency (unlike before in \autoref{fig:attack}) in DFS traversal order for the attack to succeed.}
  \label{tab:plugin_attack_success}
\end{table}

There are different plugins in Maven that can be used to package the project and its dependencies into a single uber jar.
In \autoref{tab:plugin_attack_success}, we have evaluated the attack feasibility across different Maven plugins.
The first column is the list of packaging plugins in Maven that we have tested.
The second column is the result of the attack.

\begin{lstlisting}[backgroundcolor=\color{gray!10}, caption={Class collision warning emitted by Maven Shade Plugin}, label=lst:shade,style=customstyle, numbers=none]
  [WARNING] D1-1.0.0.jar, postgresql-42.6.0.jar define 1 overlapping classes: 
  [WARNING]   - org.postgresql.Driver
\end{lstlisting}

The first two plugins, \texttt{maven-jar-plugin} and \texttt{spring-boot-maven-plugin}, lead to a successful attack based on flow described in \autoref{fig:attack}.
Next two plugins, marked with a `*', also lead to a successful attack but they emit warnings while packaging the project.
We take example of \texttt{maven-shade-plugin} to show what warnings are emitted.
As shown in \autoref{lst:shade}, multiple dependencies (D1, and postgresql) contribute with the same class name (org.postgresql.Driver) indicating potential class collisions.
This behavior does not stop and break the build and leads to successful packaging, and hence a successful attack.

\texttt{maven-bundle-plugin} leads to a successful attack but it requires to reverse the order of dependencies declared. 
In the above 4 plugins, the order of dependencies respects DFS order.
So whenever a common class is found in multiple dependencies, the one that appears first in the dependency tree is used to build the uber jar.
\texttt{maven-bundle-plugin} overrides the existing class if a new class with the same fully qualified name is found.
So we have to reverse the order of dependencies to make it work.

Finally, \texttt{maven-assembly-plugin} may or may not lead to a successful attack because it wraps the resolved dependencies into a set based on hash table.
This means that the order of dependencies is not guaranteed.
The attack is successful only if the infection dependency appears before gadget dependency when iterating over the set.
 
 \subsection{Case of Gradle}
 \texttt{\attackname} exploits specific design decisions in Maven related to packaging artifacts, classpath resolution, and the default Java class look up algorithm.
 In this section, we discuss the opportunity for the same attack on an application that builds with Gradle, an alternative build system for Java.
 
 Gradle has different design compared to Maven, regarding the construction of the dependency tree and of the final classpath. First, the classpath is generated using a breadth-first search algorithm\footnote{\url{https://docs.gradle.org/current/userguide/graph_resolution.html}}. This means that the direct dependencies are included first. This reduces the attack surface as the infection dependency must be at the same level as the gadget dependency and appear before it to be able to hijack classes from the gadget library.
 Second, custom repositories for transitive dependencies are ignored by Gradle, requiring manual declaration of the repositories in the build script of the project.
 This prevents the attacker from hiding malicious code in a self-managed repository, which might otherwise evade extra checks.
 
 The last step of the attack assumes a linear search in the classpath when loading a class. This depends on Java and is independent of Gradle or Maven.
 
 In summary, \attackname is feasible on an application that builds with Gradle. Yet, it is significantly more challenging than with Maven, due to the different ordering of the BFS algorithm and the absence of implicit download from custom repositories.

\section{Related Work}

\subsection{Dependency Conflict}

Wang et al.~\cite{wang_dependency_2018, wang_could_2019} propose the concept of dependency conflict to refer to classes of one dependency that are shadowed by the other.
Shadowing leads to a different version of a Java class being loaded.
In a more recent work of Wang et al.~\cite{wang_will_2022}, the focus is on semantic changes in methods loaded due to dependency conflict.
Cappos et al.~\cite{cappos_look_2008} analyze the security of ten widely used package managers and demonstrate that even those with cryptographic protections remain vulnerable to attacks such as replay, freeze, and extraneous dependencies, especially when malicious mirrors are involved.
Contrary to our paper, those conflicts are not adversarial.

On the mitigation front, Dietrich et al.~\cite{dietrich_dependency_2019} claim that declaring dependency using version ranges can assist with dependency conflict resolution.
However, this is not sufficient to prevent \attackname.
The malicious class can always be included in the dependency tree of the victim application, early enough in the resolution process, regardless of the versions.

Outside the Java ecosystem, Patra et al.~\cite{patra_conflictjs_2018} explore shadowing of methods in JavaScript libraries.
They report that 1 in 4 libraries may inadvertently modify or even delete the APIs of another library, causing unexpected runtime behavior and even crashes.
Wang et al. \cite{wang_smartpip_2023} propose smartPip, a system for resolving Python dependency conflicts by considering both syntactic and semantic constraints of dependencies.
Jia et al. \cite{jia_empirical_2024} conduct a large-scale empirical study on Python library dependency conflicts, finding that 60.13\% of the issues stem from interactions between third-party libraries, emphasizing how complex dependency networks can silently compromise software reliability.
Hauser \cite{hauser_hardening_2022} proposes a defense framework for mitigating dependency confusion in cloud-based CI/CD environments by leveraging a private registry's metadata and signature-based verification to detect malicious package insertions. 
Cheng et al.~\cite{cheng_conflict-aware_2022} propose PyCRE, a conflict-aware inference system that leverages knowledge graphs to reconstruct Python runtime environments, showing how semantic reasoning over third-party relationships can improve resilience to dependency misconfiguration.
Our work explores a new exploit path that leverages trusted Java dependencies by deliberately manipulating load order at runtime to manipulate the application behavior.

\subsection{Typosquatting}
Typosquatting focuses on exploiting human error in dependency names by introducing malicious packages with names similar to popular ones.
Ohm et al. \cite{ohm_backstabbers_2020} provide a comprehensive review of typosquatting attacks in the Python ecosystem, showing that even minor deviations in package names can mislead developers into using malicious code. The study highlights the stealthy and scalable nature of these attacks, especially in large ecosystems such as PyPI.
Jiang et al. present ConfuGuard~\cite{jiang_confuguard_2025}, a system that utilizes metadata and semantic embeddings to detect packet confusion attacks across six ecosystems, reducing false positives in production.
Dam et.al~\cite{dam_typosquatting_2020} conduct a large-scale analysis of typosquatting domains used for JavaScript-based scam delivery via intrusive alert boxes.
Neupane et al. \cite{neupane_beyond_2023} analyze 1,200 package confusion attacks, define 13 confusion types, and build detection tools that identify over 360,000 confusable npm package pairs missed by existing methods.

In contrast, \attackname is not based on typosquatting, but only based on dependency ordering (appears first in the classpath) to shadow legitimate classes across distinct dependencies.

\section{Conclusion}

We introduced a novel software supply chain attack, \attackname, that targets Java projects built with Maven.
We provided a proof-of-concept that illustrates the attack and demonstrated the feasibility of the attack by replicating it in a real-world project, \texttt{cwa-server}.

Our findings highlight critical vulnerabilities in the Maven and Java ecosystems that enable the \attackname attack.
The lack of control between the artifact metadata and the actual package content, combined with the class loader's behavior of loading the first matching class in the classpath without collision checks, creates a clear attack surface.
When malicious classes are introduced early in the artifact packaging order, they can override legitimate ones, especially when embedded within transitive dependencies.
Although the \texttt{Maven Enforcer Plugin} provides effective mitigation by enforcing tighter bounds and detecting conflicts at compile time, the threat remains relevant as this mitigation is not enforced by default.
This highlights the need to enforce safeguards in dependency management to protect the integrity of the Java software supply chain.
Future work may explore automated techniques for detecting dependency order tampering and evaluate the scalability of existing mitigation mechanisms, such as Maven Enforcer, in large-scale industrial projects.

\begin{acks}

We thank Herv\'e Boutemy, who is a PMC member of The Apache Software Foundation, for giving ideas about mitigations.
This work was supported by the CHAINS project funded by the Swedish Foundation for Strategic Research (SSF), as well as by the Wallenberg Autonomous Systems and Software Program (WASP), and by IVADO and the Canada First Research Excellence Fund.

\end{acks}

\Urlmuskip=0mu plus 1mu

\balance
\bibliographystyle{ACM-Reference-Format}
\bibliography{references}

\end{document}